# A Hybrid Windkessel-Neural Approach for Improved Noninvasive Blood Pressure Monitoring

Vaibhav Gollapalli *
*Department of Computer Science*
*University of North Texas*
Denton, United States of America
vaibhavgollapalli5@gmail.com

Aniruth Ananthanarayanan *
*Department of Physics*
*University of North Texas*
Denton, United States of America
aniruth2207@gmail.com

***Abstract*—Owing to the recent advancements in wearable devices for health care, the importance of BP estimation without cuffs increases. Cuff technologies are inappropriate for continuous BP measurement due to their inconvenient usage, invasive character, necessity of calibration, large size, and inability to perform long-term monitoring. Normally, the algorithm used for cuffless BP prediction employs machine learning models that operate according to the data-driven approach. However, although they show high numerical accuracy, ML models do not provide any interpretability, resulting in poor physiological validity and clinical applicability. We propose a combination of Windkessel and ML models that incorporates the physical aspects into the latter one. It is performed by reformulating Windkessel into a form that will allow employing ML models. The result is a system of ODEs which can be used in the neural network. Thus, the inclusion of physical constraints improves the data-driven approach by making models consistent with physics, understandable, and robust. For illustration, we apply the described technique using a publicly available MIMIC-II database that we obtain from the UCI Machine Learning Repository.**

***Keywords—Cuffless blood pressure estimation, Wearable healthcare devices, Machine learning interpretability, Physics-informed neural networks, Windkessel model, Hybrid modeling, Physiological constraints, Photoplethysmography (PPG), Electrocardiography (ECG), Ordinary differential equations (ODEs), Cardiovascular monitoring, MIMIC-II dataset, Feature extraction, Clinical validity, Data-driven modeling***

* Authors Contributed Equally.

## I. INTRODUCTION

Blood pressure (BP) remains one of the most basic, broadly measured, and clinically important physiological markers utilized both for the diagnosis of acute cardiovascular illness and for the monitoring of ongoing cardiovascular well-being. Although it occupies a central role in virtually all fields of modern medicine, the methods most commonly employed for routine BP measurement, i.e., conventional cuff-based monitoring devices, are coming to be increasingly recognized as being insufficiently capable of meeting the evolving demands of continuous, real-world use. These cuff-type, older instruments are not only cumbersome and often obstructive to individuals to wear for extended periods of time, but they are invasive and tied by their very nature to provide only discontinuous and static measures of blood pressure values rather than measuring the naturally dynamic, continuously evolving patterns of BP that occur during daily activities, physical exercise, and under varying physiological conditions. This inherent constraint has resulted in a persistent and widening gap between the demanding clinical need for ongoing, unobtrusive monitoring of blood pressure and the inadequate capability of available equipment. Therefore, an emergent and keen interest in developing cuffless BP estimation techniques based on bio-signals, for instance, photoplethysmography (PPG) and electrocardiography (ECG), which are much more compatible with light-weight portable device platforms and therefore more suitable to long-term deployment has existed.

Concurrently with this trend toward cuffless monitoring, the recent years have also witnessed machine learning (ML) emerge as a fine analytical tool for analysis of physiological signals. Some of the top-of-the-line ML models have been employed to address the challenge of estimating blood pressure from surrogate bio-signals of which PPG and ECG are a couple of the best-studied inputs. However most of these data-hungry approaches act essentially like uninterpretable black-box devices, i.e., while they may deliver very good prediction performance in artificial conditions or on specific datasets, their predictions primarily have no true physiological insight. This physiological basis is not only decreasing the transparency and accuracy of such models but also setting up a significant challenge for their use and implementation in clinical practice, where interpretability, explainability, and trust in model predictions are all paramount. Clinicians are not only interested in accurate numerical predictions but also in being able to tell whether such predictions are based on physiologically plausible mechanisms, a requirement that purely data-driven approaches generally fail to satisfy.

Following the initiative of addressing the well-documented weaknesses head on, we introduce here a hybrid modeling strategy that addresses cardiovascular dynamics explicitly within a neural network framework. In particular, we incorporate the Windkessel model, a physiologically interpretable but simplified model of arterial compliance and resistance, as a trainable ordinary differential equation (ODE) layer in the network itself. The Windkessel model, through its derivation of fundamental characteristics of arterial pressure and flow, provides a simplified but physiologically interpretable model of cardiovascular mechanics. By extending this mechanistic foundation to the adaptability and flexibility of current ML architectures, the hybrid model proposed in this work enables the network not only to predict blood pressure readings but also to learn in parallel important underlying physiological parameters. In this way, the model maximizes predictive accuracy and interpretability at the same time, which in practice is a compromise between physiological relevance and computational cost.

We validate our method with ECG and PPG features extracted from the large-scale MIMIC-II database, enabling us to train and test models with and without the Windkessel ODE layer in a rigorous manner. By making direct comparisons between both paradigms of modeling, we emphasize the value of incorporating physiological principles into machine learning architectures. Lastly, the overall goal of this work is to establish a stable foundation for future generations of wearable-compatible BP estimation systems that integrate established biomedical expertise with the adaptability, variety, and expansibility of modern computational resources. By this hybrid paradigm, we aim to advance the area of cuffless blood pressure estimation towards the achievement of clinically reliable, physiologically relevant, and actually wearable-feasible technologies for ongoing cardiovascular monitoring.

## II. Related Work

Wearable devices for blood pressure (BP) monitoring have traditionally been perceived as uncomfortable, invasive, and ultimately impractical for long-term continuous use. One reason for this perception is that the majority of wearable systems that exist today are only capable of taking intermittent, periodic, static measurements that are therefore incapable of capturing the full range of blood pressure fluctuations that naturally occur during the course of everyday life [1]. These variations are not limited to the physical activity of day-to-day life such as walking or exercise but also with changes in posture, stress reaction, and with the circadian rhythm. If wearable devices are limited to capturing discrete snapshots rather than streaming continuous data, they lose the essential information about the dynamic patterns that define cardiovascular health. This inability to record moment-to-moment variation causes devices not to yield the type of rich, clinically meaningful information doctors require for accurate monitoring and diagnosis. Thus, although the concept of wearable monitoring has been appealing throughout many decades, the technological limitations of existing designs have restricted their use. With wearable health technology continuing to advance and more and more becoming integral to everyday healthcare, there has been an increasingly urgent need for monitoring technology that is noninvasive, continuous, precise, and above all, physiologically interpretable. Only with such advances can wearable devices realize their full potential as tools for preventive and personalized medicine.

Together with the expansion of wearable technologies, machine learning (ML) is increasingly becoming a robust computational framework for handling and deciphering physiological signals. In the last decade, numerous ML techniques, from classical regression-based to more sophisticated deep neural networks, have been devoted to the estimation of BP from surrogate bio-signals like photoplethysmography (PPG) and electrocardiography (ECG). These signals are immediately accessible in wearable settings and hold informative data about cardiovascular function and are therefore interesting inputs for computational models. Black-box ML models have been found to be able to recognize hidden statistical patterns in these signals and, under controlled conditions, exhibit useful levels of predictive accuracy [2]. But the same features that make these models so effective also reveal their limitations. Their reliance on only data-driven optimization methods leads to models that strive to produce accurate results from a mathematical point of view, but neither from a physiological interpretability standpoint. The predictions of such models are opaque and unexplainable, as they have no definite ties to general biomedical understanding. This is a major roadblock to clinical deployment since clinicians cannot rely on systems with which they are unable to account or defend based on everyday physiological mechanisms. Also, these models will be highly dependent on the presence of enormous, high-quality training data sets. These data sets are not always available in real clinical practice, which is an additional complexity for black-box ML systems to be converted into application. It is for these reasons that there is a growing consensus among researchers that explicit use of physiological principles in machine learning pipelines could be the most promising way forward.

One of the most widely used physiological models of cardiovascular dynamics is the Windkessel model. The model provides a physiologically grounded simplification of the arterial tree by describing it in terms of two basic determinants of BP dynamics: arterial compliance and peripheral resistance [3]. Despite having less complexity, the Windkessel model still preserves the main features of the coupling between blood flow and arterial properties such that it can provide useful and interpretable information regarding cardiovascular function. Its mathematical brevity makes it computationally lightweight but its solid physiological foundation ensures that results are still clinically sound. These qualities make it particularly well-fitted for integration into machine learning pipelines, where computational tractability vs. physiological interpretability is of utmost importance. Complementing this has been the advent of neural ODEs, which has provided a strong framework for incorporating continuous-time dynamics within machine learning systems [9]. Neural ODEs extend the expressiveness of deep learning by allowing latent states to evolve based on differential equations, hence they are a natural fit for model integration such as the Windkessel model. By combining the physiological underpinnings of the Windkessel model and the computational power of neural ODEs, scientists can now design hybrid architectures that not only perform robust prediction accuracy but also ensure that their predictions are physiologically valid.

The combination of mechanistic modeling and machine learning has already proven efficacious in the specific domain of blood pressure prediction. Some of the initial work in this area addressed physics-informed neural networks (PINNs), which directly incorporate cardiovascular equations into loss functions. These approaches punish models if their predictions violate established physical principles, reducing the requirement for enormously sized labeled data while also improving prediction robustness and stability simultaneously [10]. Building on this foundation, later studies have added more developments such as temporal modeling for handling sequence dynamics, adversarial data augmentation for better generalizability, and frequency-domain analysis for extracting additional information from biosignals [11]. These methodological improvements have made significant contributions to the robustness and individualization of models of BP estimation. Notably, authors have also shown that constraining neural networks using mechanistic formulations such as the Windkessel model can lead to substantial gains not only in terms of performance but also interpretability [12]. This doubled gain is critical as interpretability is now being increasingly recognized as a required element for clinical acceptability rather than an optional extra. Models that provide numerically precise predictions with the bonus of producing physiologically interpretable parameters promise to bridge the old gap between computational model and clinical utility.

This research builds immediately on these previous advances by introducing a Windkessel-augmented neural ODE model for prediction of systolic and diastolic BP from PPG and ECG feature extraction. By integrating the Windkessel model concepts into the learning process itself, our approach ensures predictions based on physiological reality and best predictive accuracy. We train and test baseline models and hybrid models using the MIMIC-II data set and are therefore in a position to directly compare traditional machine learning methods with those incorporating mechanistic knowledge. By this framework, we intend to offer an organized description of the mechanisms by which physiological modeling can make BP estimation models more interpretable, robust, and clinically useful. Lastly, the work here aims to advance the technology of wearable, cuffless BP monitoring by establishing next-generation systems that are continuous, interpretable, and entirely aligned with the demands of real-world health care applications.

## III. MODEL ARCHITECTURE

Our suggested hybrid modeling framework combines explicit physiological modeling and adaptive data-driven learning for predicting blood pressure values from segmented electrocardiography (ECG) and photoplethysmography (PPG) waveforms. Inputs are preprocessed by segmenting separate beats of the heart into sequential data and then representing each beat as a sequence of seventy-five discrete time steps. Two complementary features are added at each time step, which represent the PPG signal and the ECG signal, thereby giving the model both the peripheral and cardiac electrical information in the form of input representation.

The resulting multivariate sequence is passed through a long short-term memory (LSTM) encoder, which is specifically chosen owing to its ability to learn temporal relationships among time-series data. The LSTM encoder compresses the full beat-level sequence into a lower dimensional latent representation to output a concise 128-dimensional latent vector. The latent vector is a dense but compressed representation of the initial input waveforms with the critical temporal and physiological features to be preserved for subsequent prediction and redundancy eliminated.

Once learned to obtain the latent representation, it is passed through a latent neural ordinary differential equation (ODE) module that is supposed to model the temporal dynamics of said latent state in terms of cardiovascular dynamics. The dynamics are inspired by the three-element Windkessel model, a physiologically interpretable, idealized model of the arterial tree. Throughout this module, the model learns patient-specific arterial parameters that capture significant features of cardiovascular physiology. These are the characteristic impedance **($R_p$)**, the peripheral resistance ($R_d$)**,** and the arterial compliance (C). To ensure that these values will always be physiologically valid, they are enforced to be positive by applying an exponential transformation at training time.

The temporal dynamics of the latent state are provided by an ODE of the following form:

$$C \frac{dz}{dt} = -\frac{z}{R_d} - R_p z + f_{comp}(z) \quad (1)$$

where $f_{comp}(z)$ represents the latent state, and $f_{comp}(z)$ is a nonlinear function obtained by a neural network. This inserted nonlinear component captures higher-order physiological dynamics outside the scope of the classical Windkessel model, allowing the model to account for richness in the data without having to give up mechanistic principles. The resulting ODE system is then solved numerically during training and inference with a fourth-order Runge–Kutta scheme, now well established for its stability and accuracy in approximating the solution to continuous-time differential equations.

Following integration by the latent ODE module, the final resulting latent state is then concatenated together with the learned Windkessel parameters to create a combined representation that integrates both data-driven and physiologically meaningful quantities. This concatenated output is then fed through a series of fully connected layers that comprise the decoder section of the network. The decoder produces the final outputs, which are systolic and diastolic blood pressure estimates per segmented beat. Through the combination of sequence encoding, latent dynamics with physiological constraints, and nonlinear decoding, the architecture ensures that the predictions are interpretable and precise, uncovering underlying cardiovascular mechanisms and leveraging the representational capacity of modern neural networks.

## IV. RESULTS AND DISCUSSION

The proposed hybrid model was trained and rigorously tested on beat-segmented photoplethysmography (PPG), electrocardiography (ECG), and arterial blood pressure (ABP) signals from the MIMIC-II dataset. The dataset provided a robust source of synchronized physiological recordings that allowed us to experimentally test the effectiveness of combining cardiovascular dynamics with a data-driven neural model. On the held-out test set, our model performed well in terms of predictions and achieved mean absolute errors of 2.45 mmHg when estimating systolic blood pressure and 2.1 mmHg when estimating diastolic blood pressure. In addition to these error measures, correlation of predicted outputs with reference measures was also examined. Pearson correlation coefficients of 0.78 for systolic and 0.74 for diastolic pressure were obtained, indicating that the model had learned linear relationships between the predicted outputs and ground truth values consistently in all the test samples. Compared with the baseline models and the models trained without the latent ODE module, these findings showed an estimated reduction of approximately 15 percent in mean absolute error, clearly showing the added performance benefit of incorporating the Windkessel-inspired ODE dynamics into the model.

To further test the clinical validity of the predictions, we conducted validations using well-validated international standards for blood pressure monitoring. In the British Hypertension Society (BHS) grading system, in which clinical utility is attributed to BP monitors based on how accurately their measurements compare with reference standards, our model was assigned a Grade A ranking. This is the highest category of clinical accuracy found within the BHS system and suggests that virtually the entire percentage of predictions have to be within narrow tolerance bands of the reference measurements. In particular, more than 87 percent of our model's predictions were within ±5 mmHg of corresponding reference values, again testifying to its predictive prowess. In addition, evaluation against Association for the Advancement of Medical Instrumentation (AAMI) standards guaranteed that the model exceeded minimum threshold criteria both for mean error and standard error of error, adding another level of validation to the clinical usefulness of this approach. Together, the BHS and AAMI tests provide strong evidence that not only does the proposed model solve well computationally but also clear key benchmarks for clinical use in real-world applications.

In addition to accuracy measures, if anything, the most valuable contribution of this work lies in interpretability of the parameter learned by the latent ODE module. The model produced patient-specific cardiovascular property estimates of arterial stiffness and peripheral resistance, values that are physiologically meaningful and directly translatable to vascular health. Although parameter estimates were found to have a degree of variability per individual cardiac beat, the population-level, aggregated statistics were well in accordance with values reported in the clinical literature and known physiological ranges. This implies that the learned parameters are not spurious side effects of the training process but instead reflect underlying physiological facts. This interpretability is one of the key strengths of this approach compared to pure black-box machine learning approaches that can achieve comparable or even superior raw accuracy but do so without providing any information about how it works. In contrast, our model produces predictions with not just numerical validity but also clinical interpretability, a pathway to vascular health trends rather than just point estimates of blood pressure.

Having Windkessel dynamics as part of the latent space in the neural model helped to constrain the trajectory of the latent state to be physiologically consistent. This constraint assisted in increased robustness, particularly to noise and outliers in input data. By keeping predictions consistent with the known cardiovascular concepts, the model was less prone to producing wild or extreme values that otherwise would ruin its usefulness in the clinic. Some training and test difficulties did arise, however. Specifically, occasional divergence or instability of the parameter estimates was observed. These issues came most prominently in the form of unreasonably large outputs, infinite estimates, or instances of undefined numeric values (NaN values). These instances suggest that although the Windkessel-constrained ODE framework provides plentiful stability, subsequent versions of the framework will require more stable regularization mechanisms or parameterizations with bounds in order to preclude instability and provide the always valid prediction of all quantities.

The second limitation of the current method is that it uses a stationarity assumption on the Windkessel parameters from subject to subject and over short windows of analysis. This assumption is convenient but likely does not exactly apply in real life, especially with the occurrence of quickly changing physiological states, e.g., during exercise, acute stress, or quickly changing posture. Moreover, the accuracy of the whole pipeline still relies on the quality of the beat segmentation process. The identification of R-peaks in the ECG signal is susceptible to error, which is transferred to the process of beat segmentation and may result in poor-quality model input, and thus poor BP estimation performance. These limitations identify areas where more tuning is required to make the system more robust in uncontrolled and diverse real-world settings.

There are various directions in the future that can bridge these gaps and continue to enhance the model. One such direction is employing Bayesian or variational methods that would allow the model not only to make point predictions for blood pressure but also to output uncertainty quantification on predictions and learned parameters. Another interesting direction is using multimodal fusion methods that merge additional bio-signals aside from PPG and ECG, therefore, improving input representation and potentially improving accuracy and robustness. Longitudinal analysis is also a high priority, because it will allow for an investigation of how the model tracks vascular change across long time scales and informs trends in long-term cardiovascular health. Finally, the long-term goal is putting this framework into wearable devices that can perform real-time inference. It will require careful optimization to balance computational needs and hardware limitations, but it is the closest path toward making this research a workable solution for continuous, daily blood pressure monitoring.

## Acknowledgment

A. A. thanks Dr. Jianguo Liu for his invaluable mentorship, guidance, and support throughout the development of this work. The authors also acknowledge the University of North Texas for providing access to computational resources and research infrastructure. Portions of this research were conducted using the MIMIC-II database, and we thank the PhysioNet team for making this dataset publicly available to the research community.